\documentstyle[12pt]{article}

\begin{document}
{\sf \begin{center} \noindent {\Large \bf Acoustic black hole evaporation as diffusion phenomena in plasmas\textbf{?}}\\[3mm]

by \\[0.3cm]

{\sl L.C. Garcia de Andrade}\\

\vspace{0.5cm} Departamento de F\'{\i}sica
Te\'orica -- IF -- Universidade do Estado do Rio de Janeiro-UERJ\\[-3mm]
Rua S\~ao Francisco Xavier, 524\\[-3mm]
Cep 20550-003, Maracan\~a, Rio de Janeiro, RJ, Brasil\\[-3mm]
Electronic mail address: garcia@dft.if.uerj.br\\[-3mm]
\vspace{2cm} {\bf Abstract}
\end{center}
\paragraph*{}
Acoustic analogues of Kerr black hole in plasmas are considered, by
taking for granted the existence of acoustic ion waves in plasmas.
An effective black holes (BH) in curved Riemannian spacetime in a
random walk plasmas is endowed with a naked singularity, when
plasmas are in the lowest diffusion mode. The plasma particle
diffusion is encoded in the effective metric. The diffusive solution
has a horizon when the plasma flow reaches the sound velocity in the
medium and a shock wave is obtained inside the slab. The sonic black
hole curved Riemannian metric is also found in terms of particle
number density in plasmas. The sonic BH singularity is found at the
center of the plasma diffusive slab from the study of the Ricci
curvature scalar for constant diffusion coefficient. It is suggested
and shown that the Hawking temperature is proportional to the plasma
Kelvin temperature through diffusion coefficient dependence to this
temperature. Therefore Unruh sonic or dumb BH is shown to have a
relation between Hawking and plasma diffusive temperatures. BH
evaporation is analogous to the diffusive phenomena in plasmas,
since in both cases Hawking temperature is inversely proportional to
mass. It is shown that Hawking analogue temperature of a plasma
torus is $T_{H}(torus)\approx{10^{-4}K}$ which is much higher than
the gravitational Hawking temperature of a one solar mass BH,
${T_{H}}^{GR}\approx{10^{-8}K}$, but still very small for being
detectable in plasma laboratories. {\bf PACS
numbers:\hfill\parbox[t]{13.5cm}{02.40.Ky: Riemannian geometries}.}}

\newpage
\newpage
 \section{Introduction}
 Relativistic and non-relativistic acoustic black holes geometries have been considered respectively by Bilic \cite{1} and Unruh \cite{2} and
 Visser \cite{3}. Other types of effective artificial black holes \cite{4} in other laboratory settings as optical and acoustic media as well as more recently
 plasma matter \cite{5} have also been addressed in the literature. These acoustic black holes are effective
 analogue pseudo-Riemannian metrics given by the homogeneous wave equation from linearised Euler flows or Navier-Stokes equation \cite{6}.
 According to Visser, the vortex plasma flow consider here
 though not irrotational can possesses a sonic BH solution. Artificial black holes have been considered with vorticity in the eikonal approximation where again homogeneous wave equation yields an acoustic BH.
 In this paper one shows that under the low diffusion limit one obtains the homogeneous wave equation from the inhomogeneous one obtained from plasma
 particle diffusion. Actually when one considers the limit of lowest diffusion mode a Riemann flat spacetime is obtained and since Riemannian curvature vanishes no
 BH is found at all. When diffusion increases a BH is found on a curved Riemannian effective plasma through the Unruh metric. Superfluid BHs have been found
 in other contexts \cite{7}. A non-diagonal sonic metric
 is obtained in this effective plasma spacetime. Actually the
 Riemannian geometry of the diffusion processes in random walks has
 been treated previously by Molchanov \cite{8}. Here the diffusion processes are shown to possibily be responsible for being the analogous to the BH evaporation, since in
 Schwarzschild and acosutic BHs the Hawking temperature are inversely proportional to the mass of the system. The presence of analogue BHs in plasmas is motivated by the acoustic ion waves which are common in non neutral plasmas.
 The paper is organised as follows: In section \textbf{2} the wave
 equation is deduced from particle diffusive equation by simply applying the
 partial time derivative in both sides of the equation. In section
 \textbf{3} a constant diffusion coefficient may lead to the presence
 of naked singularities in the case of lowest
 diffusion plasma modes in curved Riemannian effective spacetime.
 In section $\textbf{4}$ Hawking and plasma temperatures are shown to be related through the diffusion coefficient dependence of Kelvin absolute
 temperature and collision frequency of plasma particles.
 Conclusions are presented in section $\textbf{5}$.
 \newpage
\section{Acoustic BHs in effective diffusive
plasma flows} In this section we present the brief results described
above. The Riemannian acoustic effective geometry endowed with slow
diffusion can be started with the Fick's law \cite{9}
\begin{equation}
\vec{\Gamma}=n\textbf{v}=-D{\nabla}{n} \label{1}
\end{equation}
where D is the variable diffusion coefficient, and n is the plasma
particle density. The diffusion coefficient depends only on x and t
coordinates, such as $D=D(t,x)$. Thus since the diffusion
coefficient is not in principle constant $\textbf{v}$ is not
irrotational, the plasma flow is a vortex flow and the vortex
 expression becomes
\begin{equation}
\vec{\Omega}={\nabla}{\times}\textbf{v}={\nabla}D{\times}\frac{1}{n}{\nabla}n
\label{2}
\end{equation}
From the equation of particle diffusion in plasmas
\begin{equation}
{\partial}_{t}n+(\textbf{v}.{\nabla})n={\nabla}.(D{\nabla}n)
\label{3}
\end{equation}
where one has to consider that the diffusion coefficient is in
principle stationary. Actually below this would not be a problem
since one would consider only constant diffusion coefficients.
Equation (\ref{3}) yields the wave equation format required to build
the effective spacetime
\begin{equation}
{\partial}_{t}({\partial}_{t}+(\textbf{v}.{\nabla})n)={\nabla}.[(D-{\partial}_{t}D){\nabla}n-D{\nabla}({\partial}_{t}n+(\textbf{v}.{\nabla})n)=-D{\nabla}.
(\frac{D}{n}[{\nabla}n]^{2}) \label{4}
\end{equation}
which yields the wave equation in the format
\begin{equation}
{{\Box}}n=-D{\nabla}. (\frac{D}{n}[{\nabla}n]^{2}) \label{5}
\end{equation}
where the LHS of this equation is given by
\begin{equation}
{{\Box}}n=\frac{1}{\sqrt{-g}}{\partial}_{\mu}(\sqrt{-g}g^{{\mu}{\nu}}{\partial}_{\nu}n)
\label{6}
\end{equation}
where $g^{{\mu}{\nu}}$ $({\mu}=0,1,2,3)$ is the effective sonic
metric of Unruh if one of course drops the RHS of equation
(\ref{6}). But this can be easily done if one notes that that term
depends on the squared diffusion coefficient $D^{2}$. So the Unruh
sonic metric in this diffusive plasma would be reduced to
\begin{equation}
ds^{2}=(dx-(\frac{D}{n}{\partial}_{x}n+c)dt)(dx-(-\frac{1}{n}D{\partial}_{x}n+c))-dy-dz
\label{7}
\end{equation}
which represents the existence of sonic BHs in plasmas. Let us now
consider a simple example which shows that this curved BHs sonic
Riemannian metric, reduces to a Ricci-flat and Riemann-flat metric
when one considers the solution of the diffusion equation is the
lowest diffusion mode solution. To compute this example one
considers the diffusion equation above in the approximation
\begin{equation}
\frac{{\partial}^{2}n}{{\partial}t^{2}}+{\partial}_{t}[\textbf{v}.{\nabla}n]={\partial}_{t}[{\nabla}.(D{\nabla}n)]
\label{8}
\end{equation}
where ${\Omega}_{1}$ is the vorticity fluctuation according to the
rules the fields fluctuations \cite{4}
\begin{equation}
p=p_{0}+{\epsilon}p_{1}\label{9}
\end{equation}
\begin{equation}
{\psi}={\psi}_{0}+{\epsilon}{\psi}_{1}\label{10}
\end{equation}
\begin{equation}
{\rho}={\rho}_{0}+{\epsilon}{\rho}_{1}\label{11}
\end{equation}
\begin{equation}
{\vec{\Omega}}={\vec{\Omega}}_{0}+{\epsilon}{\vec{\Omega}}_{1}\label{12}
\end{equation}
Actually by noticing that diffusion equation is the generalised
conservation equation for ${\rho}=nm$ where m is the mass of
particles in plasmas one obtains the following equation
\begin{equation}
{\partial}_{t}{\rho}+{\nabla}.({\rho}{\textbf{v}})=0 \label{13}
\end{equation}
which along with the barotropic equation of state
\begin{equation}
p=p({\rho})\label{14}
\end{equation}
where p is the pressure, the generalised Unruh effective "general
relativistic" equation as
\begin{equation}
{\partial}_{t}[{c^{-2}}_{sound}{\rho}_{0}({\partial}_{t}{\psi}_{1}+\textbf{v}_{0}.{\nabla}{\psi}_{1})]
={\nabla}.[{\rho}_{0}{\nabla}{\psi}_{1}-{c^{-2}}_{sound}{\rho}_{0}\textbf{v}_{0}({\partial}_{t}{\psi}_{1}+\textbf{v}_{0}.{\nabla}{\psi}_{1})]+
{\partial}_{t}{\nabla}^{2}{\psi}_{1}+\textbf{v}_{0}.{\nabla}{\nabla}^{2}{\psi}_{1}
\label{15}
\end{equation}
from the Unruh wave equation
\begin{equation}
{\Box}{\psi}_{1}={\partial}_{t}{\nabla}.{\textbf{v}}_{1}+(\textbf{v}_{0}.{\nabla}){\nabla}.{\textbf{v}}_{1}
\label{16}
\end{equation}
Thus when one takes the hypothesis that the perturbed flow
$\textbf{v}_{1}$, besides irrotational is
incompressible
\begin{equation}
{\nabla}.{\textbf{v}_{1}}=0 \label{17}
\end{equation}
This condition of incompressibility is distinct from the used by
Visser one that it calls almost incompressible flow. Actually
insteady of the above condition one may use the fact that only the
perturbed flow $\textbf{v}_{1}$ is irrotational and that this
perturbation is on the vortex flow $\textbf{v}_{0}$ where the
diffusion coefficient is non-constant. In this case in formula
(\ref{7}) n shall be replaced by the unperturbed plasma number
particle density must be substituted by $n_{0}$. One notes that the
velocity at the horizon $g_{00}=0$ is inward velocity and therefore
the trapped surface can form. However in the next section one shall
show that naked singularity can form when the diffusion coefficient
is not constant. \section{Naked singularities in effective plasma
spacetime} R Penrose \cite{10} has investigate the presence of naked
singularities in general relativistic BHs. This is connected with a
presence of a Cosmic censorship hypothesis which forbides the
presence of naked singularities , which are essentially the
coincidence of event horizons with the physical real general
relativistic BH singularity. In this section it is shown that the
presence of a naked singularity is granted in sonic diffusive
plasmas BHs in the lowest diffusion mode where the diffusion
coefficient D is constant. The only possibility, though never been
tested, is that naked singularities may exist in the case of highly
ionised plasmas where the diffusion coefficient is nonconstant. A
lowest diffusion mode is found as a solution like \cite{9}
\begin{equation}
n=n_{0}cos{(\frac{{\pi}x}{L})}e^{-\frac{t}{\tau}} \label{18}
\end{equation}
Thus the spacetime metric (\ref{7}) becomes
\begin{equation}
ds^{2}=(dx-(\frac{D(x)}{n}{\partial}_{x}n+c)dt)(dx-(-\frac{1}{n}D(x){\partial}_{x}n+c))-dy-dz
\label{19}
\end{equation}
Thus the horizon $g_{00}=0$ surface becomes
\begin{equation}
v_{0}=D(x)tan{(\frac{{\pi}x}{L})} \label{20}
\end{equation}
which shows that at $x=0$ which a true physical singularity
coincides with the event horizon unphysical singularity. Now to
confirm that this is not a flat spacetime and that the naked
singularity is not a spurios one, one must compute the Riemann
curvature of metric (\ref{7})
\begin{equation}
R_{0101}=R_{txtx}=-DAD"+2D'A'+DA" \label{21}
\end{equation}
where $A(x)=\frac{d}{dx}ln(n^{-1})$ and the dash means derivative
with respect to the x-coordinate. Note that in at $x=0$ the Riemann
curvature component in case of constant D is
\begin{equation}
R_{0101}=R_{txtx}(t,x=0,y,z)= -D^{2}AA"|_{x=0}=
\approx{\frac{2{\pi}}{L}D^{2}}\frac{1}{tan(\frac{{\pi}x}{L})}\approx{\infty}
\label{22}
\end{equation}
which shows that the Riemannian effective plasma spacetime is curved
in the case of lowest diffusion mode and may possess a naked
singularity unless the diffusion coefficient is constant.
Confirmation of singular structure can be given by the Kretschmann
scalar as used by Royston and Gass \cite{11} in the case of optical
effective BH or as done here and in Fischer and Visser \cite{12} by
computing the Ricci curvature scalar
\begin{equation}
R=\frac{1}{A}A" \label{23}
\end{equation}
which is given by
\begin{equation}
R\approx{\frac{1}{tan(\frac{{\pi}x}{L})}}|_{x=0}\approx{\infty}
\label{24}
\end{equation}
at $x=0$. Thus the sonic BH physical singularity is at the center of
the plasma slab. Note that at the boundaries of the slabs the Ricci
scalar is given by
\begin{equation}
R\approx{\frac{1}{tan(\frac{{\pi}x}{L})}}|_{x=\frac{L}{2}}=0
\label{25}
\end{equation}
at the boundaries of the slab given by $x=\frac{L}{2}$ where L is
the width of the retangular plasma slab. Note that Roystn and Gass
\cite{11} have shown from the computation of Kretschmann scalar
\begin{equation}
R^{2}=R_{{\mu}{\nu}{\alpha}{\beta}}R^{{\mu}{\nu}{\alpha}{\beta}}\label{26}
\end{equation}
that the cosmic censorship hypothesis exists and no naked
singularity is found which does not happens here. \newpage
\section{Hawking and plasma temperatures in dumb holes}
One of the main motivations for the investigation of acoustic
analogues in effective spacetime, is the investigation of Hawking
radiation and its relation to quantum field theory. Therefore it
seems important to check for the presence of Hawking radiation in
these plasmas acoustic BHs. Let us start by computing the Hawking
temperature \cite{3}
\begin{equation}
T_{Hawking}=\frac{{h}}{4{\pi}k_{\textbf{B}}}(|\frac{{\partial}c_{sound}}{{\partial}x}|)_{c^{2}=v^{2}}
\label{27}
\end{equation}
Now let us apply this expression into the above plasma diffusive
effective spacetime solution to yield
\begin{equation}
T_{Hawking}=\frac{hD}{4{\pi}
k_{\textbf{B}}}(|\frac{{\partial}^{2}n(x)}{{\partial}x^{2}}|)
\label{28}
\end{equation}
By recalling the relation between the diffusion coefficient and the
plasma Kelvin temperature $T_{Plasma}$ \cite{9}
\begin{equation}
D=\frac{k_{\textbf{B}}T_{plasma}}{m{\nu}}\label{29}
\end{equation}
where the coefficient ${\nu}$ is the collision frequency. Thus by
substitution of (\ref{29}) into (\ref{28}) yields
\begin{equation}
T_{Hawking}=\frac{hT_{plasma}}{4{\pi}
m{\nu}}(|\frac{{\partial}^{2}ln n(x)}{{\partial}x^{2}}|) \label{30}
\end{equation}
From these last two expressions and noting that the gravitational
general relativistic Hawking temperature in Schwarzschild static BH
the Hawking temperature is inversely proportional to the mass M of
the BH as $T_{H}=\frac{h}{16{\pi}M}$. allows us to suggest that the
diffusion processes in analogue plasma effective spacetime the
diffusion phenomena would play the hole of the BH evaporation in
general relativistic BHs. Therefore a relation of type
$T_{Hawking}\approx{T_{plasma}}$ between Hawking and plasma
temperatures is obtained. Let us now consider the computation of
Hawking radiation on a more specific example of a plasma retangular
slab with a localized source \cite{9}. This can be given by the
solution of steady diffusion equation with source
\begin{equation}
{\nabla}^{2}n(\textbf{r})=-\frac{Q(\textbf{r})}{D}=\frac{{d}^{2}n(x)}{{dx}^{2}}
\label{31}
\end{equation}
where $Q(\textbf{r})$ is the source term. Due to the localization or
concentration of plasma source one obtains \cite{9}
\begin{equation}
{\nabla}^{2}n(\textbf{r})=-\frac{{\delta}(0
)}{D}=\frac{{d}^{2}n(x)}{{dx}^{2}} \label{32}
\end{equation}
which yields a Hawking temperature of
\begin{equation}
T_{Hawking}=\frac{hT_{plasma}}{4{\pi} m{\nu}}
(|\frac{d}{dx}[\frac{1}{n}\frac{d}{dx}n(x)]|) \label{33}
\end{equation}
which for a large width L of the retangular plasma slab with the
solution
\begin{equation}
{n}=n_{0}(1-\frac{|x|}{L}) \label{34}
\end{equation}
yields the Hawking temperature relation
\begin{equation}
T_{Hawking}\approx{\frac{h}{4{\pi}k_{\textbf{B}}n}}{\delta}(0)
\label{35}
\end{equation}
Of course this is not the only condition for the existence of
Hawking radiation and further investigation is necessary before one
can confirm the presence of this radiation in dumb holes analogue of
GR Kerr BHs. Let us now estimate the Hawking radiation for the plane
slab plasma geometry in (\ref{20}), by assuming that the slab is
thin and that the horizon is close to the physical singularity at
$x=0$ , thus the diffusion coefficient is
\begin{equation}
v_{0}\approx{D{(\frac{{\pi}x_{h}}{L})}} \label{36}
\end{equation}
where $x_{h}$ is the horizon locus. Since at the horizon
$v_{0}=c_{sound}$, where sound is almost trapped inside the plasma
analogue BH, thus
\begin{equation}
x_{h}=\frac{c_{sound}{L}}{{\pi}D} \label{37}
\end{equation}
since in the plasma slab one can take \cite{9} $L=\pi$
\begin{equation}
x_{h}=\frac{c_{sound}}{D} \label{38}
\end{equation}
which from the $D=4{\times}10^{3}cm^{2}s^{-1}$ yields a horizon
locus at
\begin{equation}
x_{h}= 10^{-3}c_{sound} \label{38}
\end{equation}
From this expression one finally obtains the Hawking temperature as
\begin{equation}
T_{Hawking}\approx{\frac{h}{4{\pi}k_{\textbf{B}}}}{x_{h}}\approx{10^{-11}c_{sound}}
\label{39}
\end{equation}
since the sound velocity in plasma
$c_{sound}=({\frac{{\gamma}KT_{eV}}{M}}^{\frac{1}{2}})=10^{6}KT_{eV}$
and by taking this value of $KT\approx{10}$, for a typical torus,
yields $T_{H}(Plasma)\approx{10^{-4}K}$ which still much weaker than
the astrophysical one solar mass BH Hawking temperature, where we
take a Planck constant of $h\approx{10^{-27}erg.s}$. Here one
considers that the adiabatic constant ${\gamma}=1$.

\newpage

\section{Conclusions}
 Recently a deep connection betwe
 en vortex flows and non-Riemannian artificial acoustic BHs and the
by Garcia de Andrade \cite{1}. In this report one shows that this
sonic non-Riemannian BHs is not needed in the random walk plasmas ,
even in the presence of vorticity. A naked singularity is found in
the lowest diffusion mode is obtained when the diffusion coefficient
is non-constant. In this case the Riemannian curvature does not
vanishes as when it is constant, and therefore our effective
diffusive spacetime endowed with a naked singularity is found. Note
that only large perturbations of the spherical geometry of GR BHs
may induce a naked singularity , contrary to what happens here in
the sonic black holes in plasmas in the lowest diffusion mode small
perturbation in the flow. Hawking radiation and its presence in the
dumb holes need further investigation as was done recently by
Sch\"{u}tzhold and Unruh \cite{13} in the case of the
electromagnetic wave-guides in dumb holes where quantum commutation
relations have to match as well. The interesting point here is that
in the plasma physics examples discussed the set-up for testing
Hawking radiation already exists in plasma laboratories. When a
plasma jet reaches a speed higher that the sound speed a shock wave
is formed and this is exactly the consition analogous for the
trapped sound in Unruh dumb holes in plasmas. A more detailed
investigation of the randow walk fluid motion and its connection
with the presence of acoustic BHs in the flow and Hawking BH
evaporation \cite{14} shall appear elsewhere.
\section{Acknowledgements} Thanks are due to R. Kerr and S. Bergliaffa for helpful discussions on this letter. Thanks go to Universidade do Estado do Rio de Janeiro
(UERJ) and CNPq for financial aid.

  \end{document}